\documentclass[aps,twocolumn,showpacs,prb]{revtex4-1}
\usepackage[english]{babel}
\usepackage{amsmath}
\usepackage{amssymb} 
\usepackage{times}
\usepackage{graphicx}
\usepackage{color}

\newcommand{\ave}[1]{\langle #1 \rangle}
\begin{document} 

\title{Optical response of highly excited particles in a strongly correlated system}

\author{Zala Lenar\v ci\v c$^{1}$, Denis Gole\v z$^{1}$, Janez Bon\v ca$^{1,2}$ and Peter Prelov\v sek$^{1,2}$}
\affiliation{$^1$J.\ Stefan Institute, SI-1000 Ljubljana, Slovenia}
\affiliation{$^2$Faculty of Mathematics and Physics, University of
Ljubljana, SI-1000 Ljubljana, Slovenia}

\begin{abstract}
We present a linear-response formalism for a system of correlated electrons out of equilibrium,
as relevant for the probe optical absorption in pump-probe experiments. We consider 
the time dependent optical conductivity $\sigma(\omega,t)$ and its nonequilibrium properties.
As an application we numerically study a single highly
excited charged particle in the spin background, as described within the two-dimensional  
$t$-$J$  model. Our results show that the optical sum rule approaches the 
equilibrium-like one very fast, however, the time evolution and the final asymptotic behavior of the absorption 
spectra in the finite systems considered still reveal dependence on the type of initial pump perturbation. This is observed in the evolution of its main features: the mid-infrared peak and the Drude weight.

\end{abstract}


\maketitle
\section{Introduction}
Time-resolved pump-probe optical spectroscopy represents a new powerful tool to study materials with strongly 
correlated electrons and offers a direct approach  to far-from-equilibrium phenomena, probing in particular the relaxation and thermalization processes \cite{matsuda94,okamoto2011,wall11,novelli12,dal2012,giannetti2011} . It appears that
strongly correlated systems are in general characterized very fast relaxation 
processes, emerging from the inherent strong interactions. Theoretical studies of nonequilibrium
dynamics and transient phenomena in correlated models confirm this, even in 
particular states as the Mott-Hubbard insulator \cite{eckstein11, eckstein13,lenarcic13}. 
  
In theory, probing the transient nonequilibrium state with a weak electromagnetic pulse  naturally leads to
the linear-response approach. The optical conductivity $\sigma(\omega,t)$ is the time-dependent dynamical 
quantity directly relevant to pump-probe optical spectroscopy and represents the response to the probing electric field in a  nonequilibrium situation. Some care is needed to define and properly extract the sum rules and a possible dissipationless component -- Drude weight (charge stiffness). 
Recently, such a formalism has been 
proposed for  continuous correlated systems \cite{shimizu11}. Adapted for the
application of the dynamical mean-field theory (DMFT), it has been used for the analysis of the
Hubbard model \cite{eckstein08,eckstein10,eckstein13} or with the emphasis on 
the description of time-dependent photoemission spectroscopy \cite{freericks09}. 
Another definition has been used to examine the dynamics of the Hubbard-Holstein model \cite{filippis12}. 
A stationary response, as a characteristic of a nonequilibrium quenched state, 
has also been considered recently and studied explicitly for a hard-core boson model 
\cite{rossini13} and in connection with the fluctuation-dissipation relation \cite{foini11,foini12}.
There are also studies in which the effect of the probe 
field after the intensive pulse excitation is directly followed by introduction of the (classical) 
driving time-dependent electric field \cite{moritz13}.

The aim of this paper is to introduce the differential optical conductivity
$\sigma(\omega,t)$ representing the causal linear response of the lattice current 
to an arbitrary   electric-field  pulse ${\bf E}(t'>t)$, acting on a general nonequilibrium 
many-body wave function  $|\psi(t)\rangle $ within a tight-binding model of correlated electrons. 
Such a formulation allows the definition and the consideration of the optical sum rule 
at any time, as well as the possible 
existence of the Drude weight $D(t)$ as the dissipationless response.  

As a nontrivial example we test the formalism by numerical investigation of 
a single highly excited charged particle (hole) within 
Mott-Hubbard  insulator, as represented by the two-dimensional (2D) $t$-$J$ model.  It seems plausible 
that in the long-time and thermodynamic limit, anomalous 
$\sigma(\omega,t)$ should approach 
%in a large enough system 
the ground state (g.s.) response 
$\sigma_0(\omega)$, since the particle is expected to relax to the g.s. by emitting the
extra energy via relevant bosonic excitations, i.e., via magnons. On the other hand, in 
a closed finite system one would expect the response to approach a thermal 
equilibrium response $\sigma_{th}(\omega)$, characterized by $T>0$. Our numerical 
solutions indicate that these presumptions are only partly realized for 
concrete examples and that for the long-time response also the initial state plays a role,
clearly visible at least within the limitation of our finite systems and finite evolution 
times. 
%\denis{Although this feature is observed for all numerically accessible system sizes, 
%it is unclear whether it persists in the thermodynamic limit.}

The paper is organized as follows. In Sec.~II we present the linear response formalism 
for the optical  conductivity $\sigma(\omega,t)$ within the tight-binding model 
for the general nonequilibrium many-body wave function and related density matrix.
Since the calculation of the time-dependent $\sigma(\omega,t)$ in principle involves 
a two-time evolution, implementation with a single-time evolution is developed in order to reduce the 
numerical complexity as described in Sec.~III.  Section IV is devoted to the numerical study of
a nontrivial test case, representing the optical response of the excited particle in
the strongly correlated background, as given by the $t$-$J$ model with a single hole.
Conclusions and open questions are discussed in Sec.~V.

\section{Nonequilibrium optical linear response}
In a general single-band tight-binding model of correlated electrons (with charge $e)$, 
assuming the system with periodic boundary conditions (PBCs),  the action of the electromagnetic 
field can be introduced via 
the vector potential $\mathbf{A}(t)$ through the usual gauge (Peierls) construction. 
The latter neglects the inter-band tunneling in multi-orbital models, and the field-induced 
distortions of the orbitals, but remains appropriate for the single-orbital case and weak fields.
We consider the tight-binding 
Hamiltonian (using $\hbar=1$) up to $\mathcal{O}({\bf A}^3)$,
\begin{align} \label{hat}
H(\mathbf{A}(t))&=-\sum_{i,j,s} t_{ij} \exp(ie\mathbf A(t)\cdot \mathbf R_{ij}) c_{js}^\dagger c_{is} + H_{int} \notag\\
&\approx  H_0 - e \mathbf A(t) \cdot \mathbf j + \frac{e^2}{2} \mathbf A(t) \cdot \tau \mathbf A(t),
\end{align}
written with the particle current $\mathbf{j}$ and the stress tensor $\tau$ operators,
\begin{equation} \label{jtau}
\mathbf{j}=i\sum_{i,j,s} t_{ij} \mathbf{R}_{ij} c_{js}^\dagger c_{is}, \quad
\tau=\sum_{i,j,s} t_{ij} \mathbf{R}_{ij} \otimes \mathbf{R}_{ij} c_{js}^\dagger c_{is},
\end{equation}
where $\mathbf{R}_{ij}=\mathbf{R}_{j}-\mathbf{R}_{i}$.
The electrical current 
\begin{equation}
\mathbf{j}_e(t)=-\partial H/\partial \mathbf{A}(t) =e\mathbf{j}-e^2 \tau \mathbf{A}(t)
\end{equation}
is a sum of the particle current and the diamagnetic contribution. We treat the case in which the unperturbed 
system is described by a pure (nonequilibrium) many-body wave function (wf.) $|\psi(t)\rangle$, with the time evolution  
operator $U(t',t)=\exp[-i H_0(t'-t)] $ corresponding to the time-independent $H_0$. 
We use the standard formalism \cite{nozieres64} 
to evaluate the linear response of $\langle \mathbf{j} \rangle_{t'}$ to a general ${\bf A}(t'')$ applied at $t$, 
whereas the diamagnetic part is already linear in $\mathbf{A}(t')$.
Introducing the notation for expectation values 
$\langle O \rangle_{t}=\langle \psi(t) |O| \psi(t)\rangle$ and the interaction representation 
$B^{I}(t')=U^\dagger(t',t) B U(t',t)$,
\begin{align}\label{jet}
\langle \mathbf{j}_e \rangle_{t'} &=e\langle \mathbf{j} \rangle_{t'} - e^2 \mathbf{A}(t') \langle 
\tau \rangle_{t'} +e^2\int_{t}^{t'} dt''\chi(t',t'')  \mathbf{A}(t''), \notag \\
&\chi(t',t'') =i \theta(t'-t") \langle  [\mathbf{j}^{I}(t'),\mathbf{j}^{I}(t'')] \rangle_t.
%& = i \theta(t'-t")  \langle \psi(t')| j U(t',t'') j  | \psi(t'') \rangle + c.c. .
\end{align}
Differential conductivity $\sigma(t',t)$ is defined through the response to electric field
$\mathbf{E}(\bar t)$,
\begin{equation}\label{djet}
\delta \ave{\mathbf{j}_e}_{t'}= V \int_t^{t'}d \bar t ~\sigma(t',\bar t) \mathbf{E}(\bar t) , 
\end{equation}
$V$ being the volume of the system.
Taking into account $\mathbf{A}(t^{\prime\prime})= -\int_{t}^{t^{\prime\prime}} 
\mathbf{E}(\bar t) d\bar t$ and Eqs.~(\ref{jet}),(\ref{djet}) we get
\begin{equation} \label{sigtt}
\sigma(t',t)=\frac{e^2}{V}[\ave{\tau}_{t'} -\int_{t}^{t'} dt'' \chi(t',t'') ] .
\end{equation}

For a nonstationary state there is no unique definition of the frequency-dependent $\sigma( \omega,t)$
\cite{shimizu11,eckstein13,rossini13}. We choose plausible relation reflecting the causality and switching-on of the field 
at time $t$, i.e., $\mathbf{E}(\bar t<t)=0$,  
\begin{equation} \label{sigot}
\sigma(\omega,t)=\int_0^{t_m} ds \ \sigma(t+s,t) e^{i\omega s}
\end{equation}
where $t_m$ is the width of window in which we do the Fourier transformation, so that formally $t_m \to \infty$ but is in practice the maximum time of probe duration. 
With such a definition, Eq.~(\ref{sigot}), we avoid the ambiguity of including times 
prior to the pump pulse.
The sum rule for so defined $\sigma'(\omega,t)=\mathrm{Re} \ \sigma(\omega,t)$ then
follows directly  from Eq.~(\ref{sigot}),
\begin{equation}\label{sumr}
\int_{-\infty}^{\infty}d\omega \ \sigma'(\omega,t)=\pi \sigma(t,t) = \frac{\pi e^2}{V} \ave{\tau}_t.
\end{equation}
It is evident that the sum rule, Eq.~(\ref{sumr}), is a time-dependent quantity,
i.e., $\langle \tau \rangle$ evaluated at the time $t$ when the probe field is applied.
Moreover, independent of the precise form of the Fourier transform it remains
proportional to the $\langle \tau \rangle$ at the time held fixed in the transformation.

One can define also the Drude weight $D(t)$ as the dissipationless 
component,
\begin{align}
&\sigma'(\omega,t)= 2 \pi e^2 D(t) \delta(\omega) + \sigma'_{reg}(\omega,t) \notag \\
& D(t)=\frac{1}{2V t_m}\int_{0}^{t_m} ds [\ave{\tau}_{t+s}
- \int_0^{s} ds' \chi(t+s,t+s')], \label{dt}
\end{align}
again for $t_m\rightarrow \infty$. Equation (\ref{dt}) is a generalization 
of the equilibrium expression,  $D=(1/2V)(\ave{\tau}-\chi'(\omega=0))$ \cite{jaklic00}. 
In contrast to the sum rule, $D(t)$ following from Eq.~(\ref{dt}) 
is expected to be dominated by  $t',t'' >> t$ in Eq.~(\ref{sigtt}).
Its dependence on $t$ is revealed if written in the basis of eigenstates $|\phi_m\rangle$ of $H_0$.
In the standard notation for matrix elements $\langle \phi_m| O|\phi_n\rangle=O_{mn}$ and amplitudes 
$\langle\phi_m|\psi(0)\rangle=a_m$  ($t=0$ chosen arbitrarily, e.g., when the 
nonequilibrium state is prepared),  we can express $D(t)$ in the eigenbasis, assuming that there
are no degeneracies, 
\begin{align}\label{drudebasis}
D(t)=\frac{1}{V} \sum_{m} |a_m|^2 \bigl[ \frac{\tau_{mm}}{2} - \sum_{n\neq m} 
\frac{|j_{mn}|^2}{(\epsilon_n-\epsilon_m)}\bigr] + \notag \\
+ \frac{1}{2V}\sum_{m,n\neq m} a_m^* a_n \frac{j_{mn}(j_{mm}-j_{nn})}{(\epsilon_m-\epsilon_n)} 
\ e^{i(\epsilon_m-\epsilon_n)t}.
\end{align}
Obviously, the last term provides dependence on $t$ if nonzero.
However, it is expected to vanish  if averaged over $t$ \cite{rossini13},
yielding stationary $D(t)=D_0$ which is dependent only on the (initial) nonequilibrium state $|\psi(0)\rangle$ 
through $a_{m}$. Moreover, the first two terms in Eq.~(\ref{drudebasis}) resemble the 
equilibrium expression \cite{castella95}, with thermal weights substituted by projection weights $|a_m|^2$.
However, the latter derivation is feasible only for the time-independent $H_0$.
One can express limiting $D_0$ also for the case with degeneracies, where it is of more general form,
\begin{align}
D_0=\frac{1}{V}  \sum_{\epsilon_m=\epsilon_n} a_m ^* a_n 
\bigl[ \frac{\tau_{mn}}{2} - \sum_{\epsilon_l\neq\epsilon_m} \frac{j_{ml} j_{ln}}{(\epsilon_l-\epsilon_m)}\bigr],
\end{align}
still being independent of the choice of the basis within degenerate sector.

We have so far considered the response of a pure state $|\psi(t)\rangle$ and a $t$-independent
unperturbed $H_0$.  The formalism can be extended also to more general density-matrix as well as the time 
dependent $H_0(t)$, e.g., representing the presence of the pump. 
In this case, the response to the perturbation $V(t)=f(t)V$ for an 
ensemble of pure states or time-dependent $H_0(t)$  is derived by expanding 
the density matrix up to \textit{the} first order in $f(t)$, 
$\rho(t)=\rho_0(t) + \rho_1(t) + O(f^2)$, and using the 
von-Neumann equation  $\partial \rho(t)/\partial t =-i~[H_0(t)+ f(t)V,\rho(t)]$. 
The linear response of general operator $B$ at time $t'$ to perturbation applied at $t$ is then
\begin{align}
\delta \ave{B}_{t'}&=Tr[\rho_1(t') B] \\
&=-i \int_{t}^{t'} dt'' f(t'')Tr[\rho_0(t) [B^{I}(t'),V^{I}(t'')]]
, \\
\rho_0(t)&=U(t,0)\rho_0(0)U^\dagger(t,0), \notag \\
\rho_1(t')&=i \int_t^{t'} dt'' f(t'')U(t',t'') [\rho_0(t''),V] U^\dagger(t',t''), \notag
\end{align}
where the time-evolution operator is $U(t',t)=\hat T[\exp(-i\int_t^{t'}H_0(t'') dt'')]$. 
In such a formulation
 %the third term in Eq. (\ref{jet}) that stands for 
 the linear response of the particle current to the field applied at $t$, written with density matrix for a single pure state $\rho_0(t)=|\psi(t)\rangle \langle \psi(t)|$ is
\begin{align}
\delta \langle \mathbf{j} \rangle_{t'}=-i \int_{t}^{t'} dt'' (-e \mathbf{A}(t''))
Tr[\rho_0 (t) [\mathbf{j}^{I}(t'), \mathbf{j}^{I} (t'')].
\end{align}
Optical conductivity, possibly generalized also to an ensemble of pure states, is then
\begin{equation} \label{sigttgen}
\sigma(t',t)=\frac{e^2}{V}\left(\ave{\tau}_{t'} -
i\int_{t}^{t'} dt'' Tr[\rho_0 (t) [\mathbf{j}^{I}(t'), \mathbf{j}^{I} (t'')]]\right),
\end{equation}
where now $\ave{\tau}_t=Tr[\rho_0(t) \tau]$.

\section{Numerical implementation}

Let us discuss here the numerical implementation for $\sigma(\omega,t)$ 
only for the case of single wf.  $|\psi(t)\rangle$ and time-independent $H_0$.  
The apparent disadvantage of the definition with Eq.~(\ref{sigtt}) is that for a fixed $t$ the evaluation 
of $\chi(t',t'')$ via Eq.~(\ref{jet}) requires the propagation $U(t',t'')$ for each $t''$, and finally for the 
numerical calculation at chosen $t_m$, $\mathcal{O}(t_m^2/\Delta t^2)$ operations are needed with
$\Delta t$ being the integration time step.
 
Instead it is more efficient to calculate the integral inside the matrix element, Eq.(\ref{jet}),
performing discrete steps,
\begin{align}
\int_t^{t'} dt''U(t',t'') j|\psi(t'')\rangle  
\approx\Delta t \sum_{n=0}^{n_m}U_\Delta^{n_m-n+1} \ j|\psi(t+n\Delta t)\rangle, \label{intsum}
\end{align}
where $n_m=(t'-t)/\Delta t -1$ and $U_\Delta=U(t''+\Delta t,t'')=\tilde U(\Delta t)$
propagation for $\Delta t$. The sum (\ref{intsum}) 
can be then evaluated recursively,
\begin{align} 
|S_0\rangle=j|\psi(t)\rangle, \
|S_{n}\rangle=j|\psi(t+n\Delta t)\rangle + U_\Delta|S_{n-1}\rangle \label{iter}
\end{align}
so that finally
\begin{equation}\label{sigiter}
\sigma(t',t)=(e^2/V)(\ave{\tau}_{t'} + 2 \Delta t \ \mathrm{Im} \langle \psi(t')|j \ \tilde{U}(\Delta t)|S_{n_m}\rangle).
\end{equation}
Such a procedure reduces the number of operations to $\mathcal{O}(t_m/\Delta t)$. 
We note that quite an analogous procedure can be applied for other transient 
correlation functions or for the time-dependent $H_0(t)$.

\section{Single excited particle}

In order to test the feasibility of the above formalism and contribute to the discussion of 
transient optical response of nonequilibrium strongly correlated 
systems, we investigate in the following the case of a single excited 
charge carrier (hole) $N_h=1$, doped in an antiferromagnetic Mott-Hubbard insulator. 
We consider the standard single-band $t$-$J$ model, 
\begin{equation} \label{tj}
H_{0}=- t_h \sum_{\langle i,j\rangle,s} (\tilde{c}_{i,s}^\dagger \tilde{c}_{j,s}  + \textrm{H.c.}) + 
J\sum_{\langle i,j\rangle} (\mathbf{S}_{i} \cdot \mathbf{S}_{j} -\frac{1}{4} n_i n_j ),
\end{equation}
where $\tilde{c}_{i,s}=c_{i,s}(1-n_{i,-s})$ are fermion operators, projected onto the space 
with no double occupancy, describing hopping between the nearest neighbor sites only. 
We consider in the concrete example
the 2D square lattice,  relevant for cuprates, with $J/t_h=0.4$.
Further on we use $t_h=1$ as the unit of energy, as well of time $t_0=\hbar/t_h=1$.
Lattice spacing is set $a_0=1$ so that $V=N$, as well $e=1$.

The intention is to consider the situation 
relevant for the pump-probe experiments on cuprates \cite{matsuda94,okamoto2011}. One can
imagine two different situations: 

\noindent a)  First is the photodoping of the Mott-Hubbard insulator,
where a low concentration of highly excited charge carriers (holons and doublons) is created 
within an otherwise insulating AFM system. Both types of carriers would exhibit, at least in the transient stage,
an independent response to the  probe pulse. To simulate this situation we consider as the initial
wf. a single hole  localized on one site, $|\psi(0)\rangle $ being the eigenstate of 
the model, Eq.~(\ref{tj}), where effective $t_h$ is put to zero. 

\noindent b) Another setting is a weakly doped AFM insulator, represented by
the g.s. of a single hole within the $t$-$J$ model. 
The effect of a strong pump pulse applied to it can be simulated by introduction of a phase shift in the hopping term of Eq.~(\ref{tj}), 
changing $t_h \to t_{h}^{ij}=t_h {\exp}(i \theta_{ij}) $, where we perform the maximum shift in the chosen
direction $x$, i.e. $\theta_x =\pi$. Both scenarios correspond to the same change
in the total kinetic energy $E_{kin}$, calculated from the expectation value of the $t_h$ term in Eq.~(\ref{tj}).
As obvious for the initially localized hole with $t_h(t=0)=0$, the excited
state has $E_{kin}(t=0)=0$. For the choice $\theta_x=\pi$ the effect of
particular phase shift is to change the sign of hopping in the $x$ direction. 
Due to rotational invariance this again yields zero total kinetic energy.

Results for a single hole within the $t$-$J$ model are obtained via two numerical methods. 
One is the exact diagonalization (ED) of small systems employing Lanczos method, where we study the 2D square lattices 
with  $N \leq 26$ sites and p.b.c. First we find the g.s. $|\psi(0)\rangle$.
By solving time-dependent Schr\"odinger equation, time evolution of $|\psi(t') \rangle$
and evaluation of recursive relations, Eq.~(\ref{iter}),
is obtained by employing the Lanczos basis \cite{park86,prelovsek13}. 
Since the available square lattices ($N =18,20, 26$) are in general not rotationally invariant 
we perform the averaging of 
$\sigma'=(\sigma'_{xx}+\sigma'_{yy})/2$ for the case b) together with 
corresponding pulses $\theta_{\alpha}= \theta_{x}, \theta_{y}$.

Another method to evaluate $\sigma(\omega,t)$ is the diagonalization 
within the limited functional space (EDLFS) \cite{bonca2007,prelovsek13}   
The advantage of the EDLFS method 
in the equilibrium regime follows from a systematical construction of states 
with distinct  configurations of local spin excitations in the proximity of the hole.
In this way (in contrast to the ED on small system) the method in 
principle deals with an infinite system. In practice the effective size of the 
system is larger than in ED, but still limited by the number of basis states taken
into account.  The EDLFS remains efficient even when applied to nonequilibrium  
systems,
as long as the spin disturbance 
caused by the local quench remains within the confines  of  generated  
spin  excitations \cite{mierzejewski11,vidmar11}. In the considered case 
spin excitations extend up to $L=16$ lattice sites away from 
the hole.

First we analyze the time variation of $\langle \tau_{\alpha\alpha} \rangle$, representing the sum rule
Eq.~(\ref{sumr}), which is for the tight-binding model (\ref{tj}) related to the kinetic energy, 
$\langle \tau_{\alpha\alpha} \rangle_t = -E_{kin,\alpha}(t)$, where only hopping in 
$\alpha$ direction is taken into account.  In Fig.~1 we present ED and EDLFS results for 
$\langle \tau \rangle$ calculated directly from kinetic energy for 
$|\psi(0)\rangle$  corresponding to localised and $\pi$-pulse excitations, respectively.
Averaging $\langle \tau \rangle=(\langle \tau_{xx} \rangle+\langle \tau_{yy} \rangle)/2$
is employed in ED calculations, whereas for EDLFS $\langle \tau \rangle = \langle \tau_{xx} \rangle$.
For comparison we show also the g.s. $E_{kin,x}^0=E_{kin}^0/2 \sim - 1.4$. From Fig.~1
it follows that for both types of  initial excited states the decay of $\langle \tau \rangle$
 is very fast. The corresponding short time $t_d$ can be related to the formation
 of the spin polaron  and can be explained with the generation of string states \cite{golez13}.
 It is expected to scale as $t_d \propto (J/t_h)^{-2/3}$. 
 
We confirm in Fig.~1 that both methods, the ED and the EDLFS, give quite consistent result for 
$\langle \tau\rangle_t$ for short times $t<t_d \sim 1.5$ (for chosen $J = 0.4$). 
For intermediate times  $t_d<t < t_i \sim 15 $ the decay of $E_{kin,x}^0$ 
evaluated with ED is somewhat slower, which could indicate the influence of finite-size 
and p.b.c. effects. Namely, in small systems considered with the ED spin excitations populate 
the lattice and consequently influence further hole relaxation.
The spread of spin excitation is expected to saturate in $t_i \sim \sqrt{N}/J$,
yielding approximately stationary response afterwards.
Within the  EDLFS with in principle an infinite lattice, such effects are not present or appear 
only at later times due to the restricted basis. 
More than an artifact, ED results on small lattices should be relevant 
for the optical sum rules of systems with finite density of carriers,
while those from EDLFS corresponds to systems with vanishing density. 

Although for both types of excitations 
$E_{kin}(t=0) \sim 0$, the distinction in sum rules is apparent for times shorter than $t<t_d$,
featuring the fact that the state of initially localized hole has the rotational symmetry, 
whereas with $\pi-$pulse this symmetry 
is broken, with $E_{kin,x}(t=0)=-E_{kin,y}(t=0)$. 

\begin{figure}[ht]
\begin{center}
\includegraphics[width=0.31\textwidth]{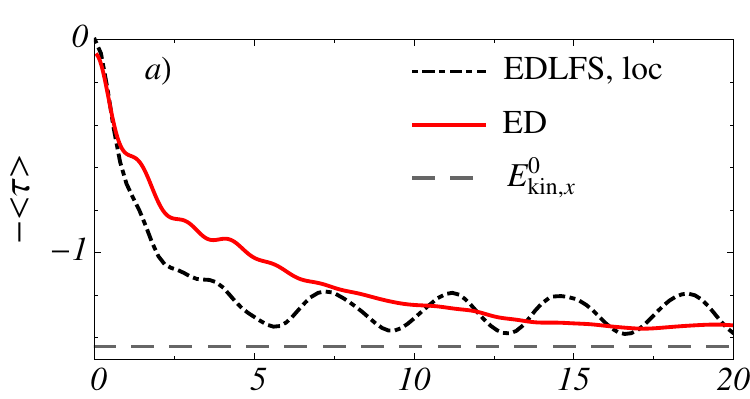}\\
\includegraphics[width=0.31\textwidth]{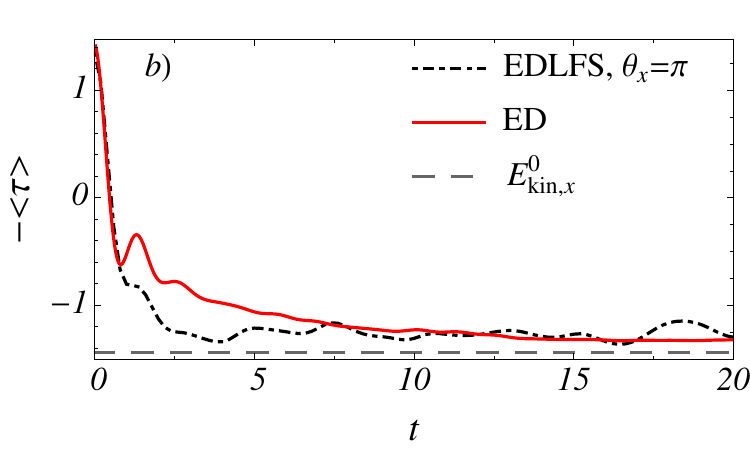}
\end{center}
\caption{(Color online) Sum rule $\langle \tau \rangle$ vs. time $t$ for $J=0.4$ as obtained using the 
ED  on $N=26$ sites as well as the EDLFS. Results are presented for initial states of: 
a) the localized hole and b) the $\pi$-pulse. The g.s. $E^0_{kin,x}$ value is also displayed.}
\label{Fig1}
\end{figure} 

\begin{figure}[b!]
\begin{center}
\includegraphics[width=0.32\textwidth]{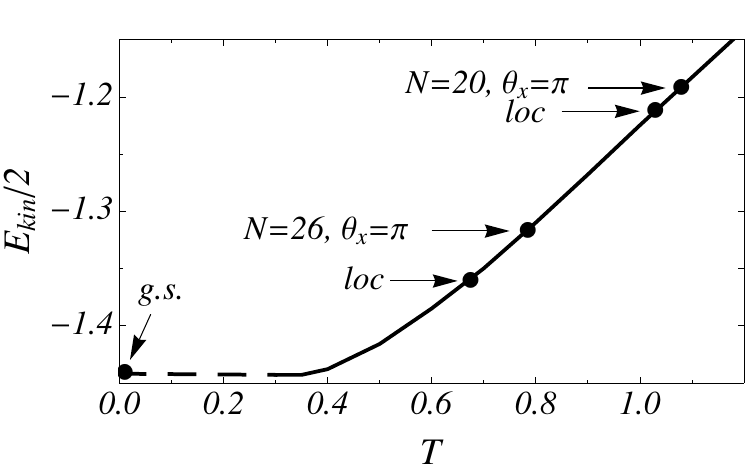}
\end{center}
\caption{Equilibrium kinetic energy $E_{kin}/2$ vs. temperature $T$ for single hole on a system with
$N=26$ sites. Dots mark $E_{kin}(t\rightarrow \infty)/2$ for both types of excitations in $N=20, 26$ systems and g.s..}
\label{Fig1a}
\end{figure}

It is evident that for long times $t>t_i$
kinetic energy approaches or oscillates around a quasi-stationary value, denoted 
by $E_{kin}(t\rightarrow \infty)$. Within the ED latter still somewhat depends on the 
system size $N$. The larger is the system under consideration, closer is this value to 
the g.s. $E_{kin}^0$, as marked in Fig.~\ref{Fig1a}.

Since the ED simulates a fixed-size system, and the excited state is quite far from the g.s., 
a plausible  interpretation could be investigated within the concept of thermalization, i.e. the approach 
to the equilibrium state with a finite effective temperature $T_{eff} >0$. 

One could argue that 
different $E_{kin}(t\rightarrow \infty)$ originate in different
$T_{eff}$, depending on the type of quench and size of the system.
In a finite system $T_{eff}$ is set by the excitation energy so that the canonical expectation value of 
the energy equals the total initial energy $\langle \psi(0)|H_0|\psi(0)\rangle=E_{tot}$. 
At $N=26$ the effective temperature is approximately $T_{eff}=0.35,0.5$ 
for the initially localized hole and $\pi$-pulsed hole, respectively, and increases as $N$ decreases.

Our observation is that $E_{kin}(T_{eff})$ is still lower than 
$E_{kin}(t\rightarrow \infty)$ for all finite systems considered, as seen in Fig.~\ref{Fig1a}
This suggests that system cannot 
completely thermalize, possibly due to the discreteness of spin excitations in finite systems.
In this connection we notice that $T>0$ calculation of the  same model, Eq.~(\ref{tj}), 
with a single hole using the finite temperature Lanczos method \cite{prelovsek13,zemljic05}
surprisingly reveal that $E_{kin}(T\sim J)$ does not essentially differ from g.s. $E_{kin}^0$, Fig.~\ref{Fig1a}.
%reveal any $T$-dependence (in particular even slight decrease) for $T<  0.4 =J$. 
 
\begin{figure}[b!]
\begin{center}
\includegraphics[width=0.31\textwidth]{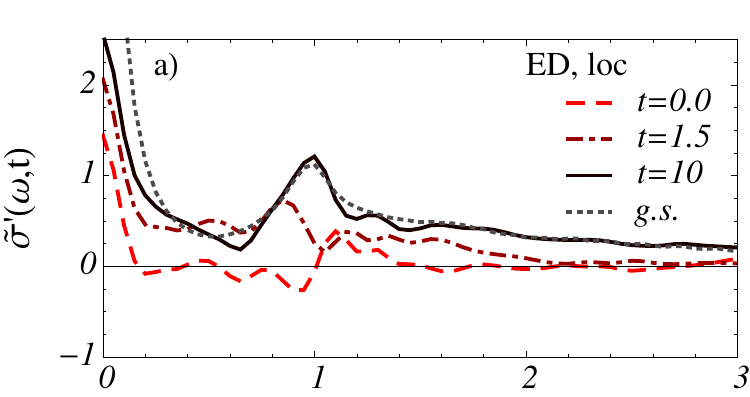}\\
\includegraphics[width=0.31\textwidth]{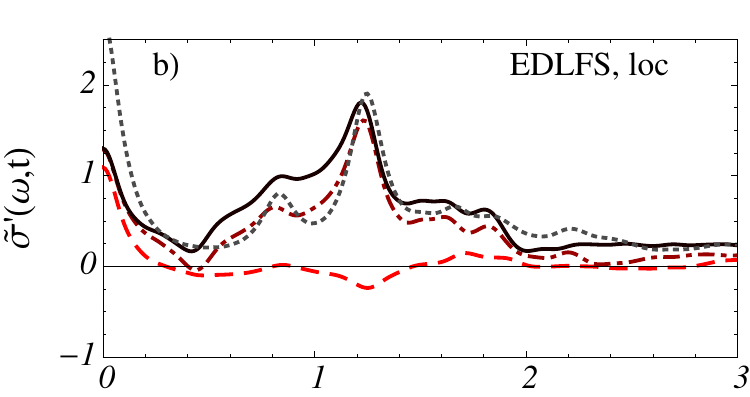}\\
\includegraphics[width=0.31\textwidth]{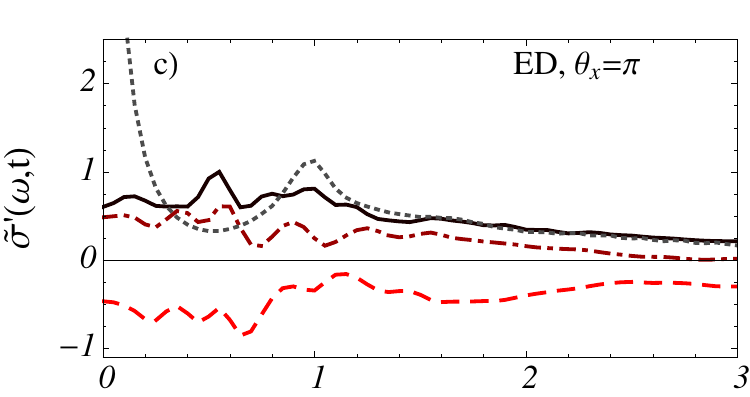}\\
\includegraphics[width=0.31\textwidth]{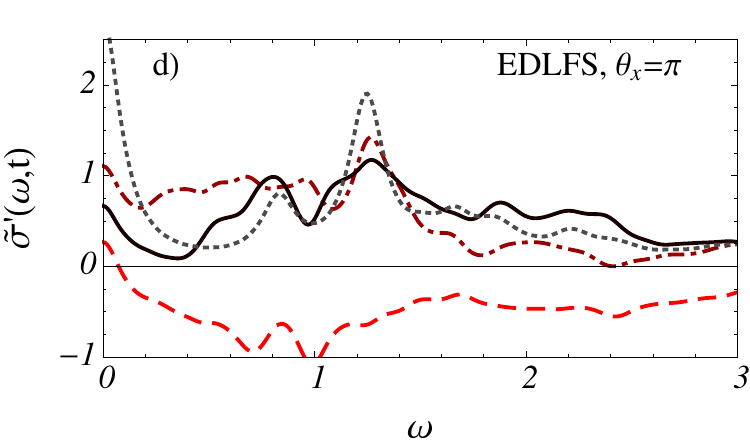}
\end{center}
\caption{(Color online) Time-dependent optical conductivity $\tilde\sigma'(\omega,t)$ (per hole)
vs. $\omega$  as calculated at different times 
$t = 0, 1.5, 10$ for: a) the initially localized hole with the ED on $N=26$ sites, b) localized hole with 
with the EDLFS, c) the $\pi$-pulse excited hole with the ED, and d) the $\pi$-pulse
excited hole within EDLFS. For comparison the g.s. $\tilde\sigma_0(\omega)$ is shown. 
Broadening of spectra $\delta \omega=0.1$ is used.
}
\label{Fig2}
\end{figure}

\begin{figure}[b!]
\begin{center}
\includegraphics[width=0.31\textwidth]{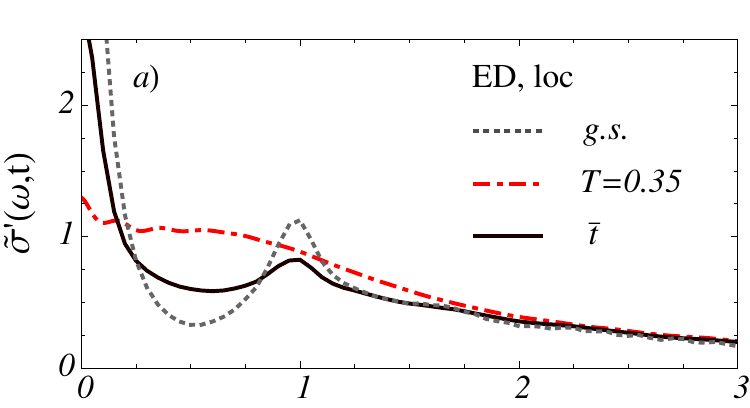}
\includegraphics[width=0.31\textwidth]{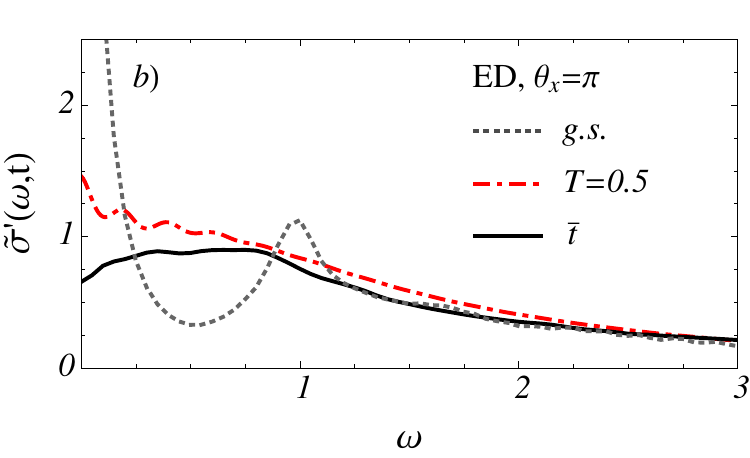}
\end{center}
\caption{\label{Fig3}(Color online) Long-time response $\tilde\sigma'(\omega,\bar t)$ vs. $\omega$ 
as calculated with ED on $N=26$, obtained by average over responses in interval $t\in[20,60] \gg t_d$ for a) initially 
localized hole, and b) the $\pi$-pulse excited hole. Results are compared
to the g.s. $\tilde\sigma_0(\omega)$ and the thermal $\tilde\sigma_{th}(\omega)$ at corresponding effective temperatures
$T=0.35,0.5$. Spectra are broadened with $\delta \omega=0.1$. }
\end{figure}

In Fig.~\ref{Fig2} we finally present results for the time-dependent optical 
spectra per hole $\tilde{\sigma}'(\omega,t)=N\sigma'(\omega,t)$
as obtained by ED and EDLFS, respectively, following the described procedure.
The cases of initially localized and $\pi$-pulse excited holes are compared. With respect
to Fig.~1 chosen times represent different evolution stages: 
a) $t \sim 0$ response of the initial excited state $|\psi(0)\rangle$, 
b) response at approximately characteristic decay time $t  =1.5  \sim t_d$, and c) 
$t=10 \sim t_i$ already relaxed but 
not yet fully  stationary response. 
To mimic the stationary response for finite systems $\tilde{\sigma}'(\omega,\bar{t})$, obtained by
average over responses in interval $t\in[20,60] \gg t_d$, is presented in Fig.~\ref{Fig3} and compared with the g.s. $\tilde\sigma_0(\omega)$ 
and with the thermal-equilibrium result $\tilde\sigma_{th}(\omega)$ at 
effective $T_{eff} \gtrsim J$, all obtained within the ED. 

In the initial stage the response is very incoherent and for the $\pi-$pulse case even predominantly negative 
$\tilde{\sigma}'(\omega,t) <0$, which is compatible with the sum rule $\langle \tau \rangle_{t\sim 0} < 0$ in 
Fig.~1. The latter indicates on highly nonequilibrium state $|\psi(0)\rangle$ corresponding to an inverse particle
population \cite{naoto09}. 

With some quantitative difference between both methods, 
within the g.s. $\tilde\sigma_0(\omega)$ two features are well visible and of particular interest. 
One is the mid-infrared (MIR) peak at $\omega \sim 2.4~J$, which
is the signature of the string-like excited particle states within the 2D AFM background.
We see in Figs.~\ref{Fig2}a,b  that the MIR-like peak appears also in $\tilde\sigma(\omega,t)$ and 
stabilizes very fast  at $t \sim t_d$ for the localized hole, independent of the numerical 
method employed.
On the other hand, after the $\pi$-pulse (Fig.~\ref{Fig2}c,d) the MIR peak is less pronounced. 
Especially within the ED, $\tilde\sigma(\omega,t>t_d)$ is closer to the thermal 
$\tilde\sigma_{th}(\omega)$ at appropriate $T_{eff} = 0.5$.
In the equilibrium at such high $T>J$ the MIR peak (Figs.~\ref{Fig2}c, \ref{Fig3}b) is already 
smeared out due to thermally disordered AFM spin background. 
Within the EDLFS results (Fig.~\ref{Fig2}d) the MIR peak recovers better, which could be attributed to 
a lower density of spin excitations in effectively bigger systems.

The second feature, pronounced within the g.s. is the Drude weight $D$, i.e. peak at $\omega=0$ 
which accounts for $\sim 1/3$ of the weight in the sum rule at $T=0$. It should acquire a 
finite width in the equilibrium at $T>0$  due to scattering processes (i.e. in the strict sense 
$D=0$ is expected for a nonintegrable system).  Although within both methods used 
(Eq.~\ref{drudebasis} would require full ED method) we cannot strictly establish and determine 
the value of $D(t)$ as defined in 
Eqs.~(\ref{dt},\ref{drudebasis}), it is evident that with the respect of low-$\omega$ response  
studied, excited particles reveal quite different behavior. Within the ED $\pi$-pulse excited hole 
shows  essentially no Drude peak, i.e., no corresponding low-$\omega$ remainder
at $t>0$. On the other  hand, initially localized hole displays a substantial low-$\omega$ peak
and presumably $D(t)>0$  (note that we use broadening $\delta \omega = 0.1$) 
at all $t>0$, although  the weight is smaller than in $\tilde\sigma_0(\omega)$. 
Using the EDLFS we notice a weak remainder of the low-$\omega$ contribution even for 
$\pi$-pulse, yet much  smaller than for the localized hole. This findings indirectly support 
Eq. (\ref{drudebasis}), that also the initial wf. (and not just its total energy) determines the 
limiting Drude weight value $D_0$.

\section{Conclusions}
We have presented a formalism for the linear optical conductivity response $\sigma(\omega,t)$ 
of a nonequilibrium (excited) state of  a strongly correlated system, 
where probe pulse is taken as a perturbation in the linear order.
 In the absence of an unique approach we have 
chosen the definition reflecting the onset of the probe electric field pulse
at time $t$, which is in contrast to some other studies. Such an approach allows the
discussion of the optical sum rule as well as the dissipationless Drude weight at any 
time $t>0$. On the other hand, the definition introduces a complication due to an 
additional time integration which we circumvent by a particular numerical
 implementation.   

The presented test case of a single highly excited charge carrier within the $t$-$J$ model 
already shows several features and opens questions relevant for the theoretical analysis of 
the pump-probe spectroscopy results. 
Independently of the initial state we observe a fast relaxation of several 
observables, e.g. the kinetic energy and the optical sum rule, 
towards the respective g.s. value. Still, more specific
features of the transient and long-time optical response, as the MIR peak and
the Drude component $D(t)$,  appear non-universal. Our results reveal  
that they do not depend  merely on the excitation energy,  as e.g. expected from the 
canonical thermalization, but as well on the character and wf. of the initial excited state. 
The persistance of this feature in the thermodynamic limit remains an open question.
One one hand, it appears plausible that in an infinite system the local state of the quasiparticle 
will correspond to the ground state. Nevertheless, this statement is, e.g., not evident 
for a quantum system with gapped bosonic excitations.
To this end our findings show lack of canonical thermalization, observed before in 
theoretical \cite{rigol08,manmana07} as well as experimental studies \cite{giannetti09}. 
Since we address excited systems, the concepts of thermalization and relaxation
to the g.s. response are only partly applicable, especially for dynamical quantities, and 
remain the challenge also for further studies. We should note as well that in the application 
of our formalism and results to the pump-probe experiments some care is needed when 
energy absorption of particular probe pulses is measures, which cannot
be directly compared to time-dependent $\sigma'(\omega,t)$ calculated in the
present study. %\cite{schins07}.

Z. L. and D. G. acknowledge discussions with M. Eckstein and A. Silva. J. B. acknowledges stimulating  discussion  with S. A. Trugman and the financial support by the Center for Integrated Nanotechnologies, an Office of Science User Facility operated for the U.S. This work has been supported by the Program P1-0044 and the project J1-4244 
of the Slovenian Research Agency (ARRS).

\end{document}